\documentclass[aps,pra,twocolumn,superscriptaddress,floatfix]{revtex4-1}

\usepackage{graphicx}
\usepackage{amsmath, amsfonts, amssymb, bbold, bm, commath,enumitem}
\usepackage{color}
\usepackage{wasysym}
\usepackage{subcaption}
\usepackage{multirow}
\usepackage{epstopdf}
\usepackage{natbib}
\usepackage{braket}

\usepackage{url}
\usepackage{soul}

\newcommand{\up}{\uparrow} 
\newcommand{\down}{\downarrow} 
\newcommand{\Order}{\mathcal{O}}
\newcommand{\I}{\mathcal{I}}

\newcommand{\Hami}{\mathcal{H}}
\newcommand{\TR}{\mathcal{T}}
\newcommand{\PH}{\mathcal{C}}
\newcommand{\K}{\mathcal{K}}

\newcommand{\hp}[4]{ \begin{pmatrix}
  #1 & #2 \\
  #3 & #4 
\end{pmatrix}}
\newcommand{\vp}[2]{ \begin{pmatrix}
  #1 \\
  #2
\end{pmatrix}}

\def\eq#1{Eq.~(\ref{eq:#1})}
\def\fig#1{Fig.(\ref{fig:#1})}
\def\sfig#1{(\subref{fig:#1})}

\begin{document}

\title{Nonlinear optical effects in inversion-symmetry-breaking superconductors}
\author{Tianrui Xu}
	\affiliation{Department of Physics, University of California, Berkeley, California 94720, USA}
	\affiliation{Materials Sciences Division, Lawrence Berkeley National Laboratory, Berkeley, California 94720, USA}
\author{Takahiro Morimoto}
	\affiliation{Department of Applied Physics, The University of Tokyo, Hongo, Tokyo, 113-8656, Japan}
\author{Joel E. Moore}
	\affiliation{Department of Physics, University of California, Berkeley, California 94720, USA}
	\affiliation{Materials Sciences Division, Lawrence Berkeley National Laboratory, Berkeley, California 94720, USA}

\date{\today}

\begin{abstract}
We study nonlinear optical responses in superconducting systems with inversion ($\I$) symmetry-breaking order parameters. 
We first show that any superconducting system with $\I$ and time-reversal ($\TR$) symmetries requires an $\I$-breaking order parameter to support optical transitions between particle-hole pair bands. 
We then use a 1D toy model of an $\I$-breaking superconductor to numerically calculate linear and nonlinear conductivities, including shift current and second harmonic generations (SHG) responses. 
We find that the magnitude of the signal is significantly larger in shift current/SHG response compare to the linear response due to the matrix element effect.
We also present various scaling behaviors of the SHG signal, which may be relevant to the recent experimental observation of SHG in cuprates~\cite{zhao16natphys}. 
Finally, we confirm the generality of our observations regarding nonlinear responses of $\I$-breaking superconductors, by analyzing other models including a 1D three-band model and 2D square lattice model. 
\end{abstract}

\maketitle


{\it Introduction.--} A wide variety of optical measurements are used to give insight into complex ordered states of quantum materials~\cite{orenstein_ultrafast,giannetti16}.  Measurements of linear optical conductivity in cuprate superconductors, for example, have been used to probe spectral weight transfer~\cite{vandermarel} and universal dissipation rates~\cite{orenstein_planckian}.  Optical measurements beyond the linear regime can be divided between conventional nonlinear optics~\cite{Boyd}, which can be analyzed theoretically using perturbation theory and is the subject of the present work, and far-from-equilibrium pump-probe measurements~\cite{kampfrath13,aoki14rmp}.

Recently, there have been experimental efforts on non-linear optical effects of cuprates in pseudogap phases~\cite{zhao16natphys,zhao18chpt}.  Such experiments have shown that the symmetry of quantum phases, or lack thereof, can be accessed by non-linear optical measurements, because second-order optical effects such as photocurrent and second-harmonic generation require broken inversion ($\I$) symmetry.  A theoretical analysis was recently carried out of {\it third-order} nonlinear response~\cite{benfatto}; the third-order susceptibility is generally nonzero in all solids but its frequency dependence can probe excitations such as the Higgs mode of a superconductor~\cite{matsunaga12,matsunaga13,matsunaga14}.

In this Letter, we study how $\I$ symmetry breaking in the superconducting state affects linear and non-linear optical conductivities (shift current and second harmonic generations (SHG)~\cite{sipe00prb,rappe12,cook17natcomm,morimoto16,patankar18}).  The non-linear consequences of $\I$-breaking in such systems are stronger and onset more rapidly near the superconducting transition than the linear ones. To show this, we consider a superconducting system described by Bogoliubov-de Gennes (BdG) formalism, with $\I$-symmetry-preserved electronic structure.  We first review how if such a system preserves both time-reversal ($\TR$) symmetry and $\I$ symmetry, there will be no first- and second-order optical transitions between particle-hole bands at the gap energy, then show how such transitions become possible with broken $\I$. 

We then show numerical simulations of the linear and non-linear optical conductivities for a 1D toy model of $\I$ breaking superconductors. For simplicity, we consider singlet pairing terms in our models and assume that $\TR$ remains a good symmetry; the (linear) Kerr effect created in a $\TR$-breaking superconductor has been actively studied for many years~\cite{kapitulnik}. We first consider a 1D chain with alternating hopping (essentially the Su-Schrieffer-Heeger model of polyacetylene~\cite{ssh80}) and an $\I$-breaking superconducting order parameter. We show how optical conductivities scale with $\I$-breaking, and find that the scaling is stronger for non-linear optical conductivites when $\I$-breaking is small. We also show the temperature dependence of SHG signals at experimentally accessible region, which is consistent with recent experimental observations~\cite{zhao16natphys}.

In closing, because tight-binding models with two bands can be misleading or special in some cases because the upper band is essentially determined by orthogonality to the lower band, we show that the non-linear optical effects from $\I$-breaking persist in other models such as a 1D three-band model and a 2D minimal model of the cuprate band structure (see Supplemental Material~\ref{app2D}).


{\it Model and conductivities.---} 
We consider a superconducting system described by the Bogoliubov-de Gennes (BdG) formalism:
\begin{align}
    \Hami=\frac{1}{2}\sum_k\Psi_k^\dagger\hp{H(k)}{\Delta}{\Delta^*}{-H^T(-k)}\Psi_k,
    \label{eq:bdg}
\end{align}
where $H(k)$ models the electronic structure, $\Delta$ is the superconducting pairing order parameter, and $\Psi_k$ is a Nambu spinor wavefunction~\cite{nambu60pr,toposcrev17,dgscbook,schriefferscbook,fu08}. 
The couplings to external electric fields are introduced by the minimal coupling prescription: $H(k)\to H(k+eA)$. 
In the expansion of $H(k+eA)$ with respect to $A$, the second- and third-order terms in $A$ are called paramagnetic and diamagnetic currents, and are relevant to the linear and second-order non-linear responses, respectively. Specifically, one defines the velocity operator and its derivative:
\begin{align}
    \hat v^a(k)&=\partial_{k_a}H(k)\tau_0, &
    \hat w^{ab}(k)&=\partial_{k_b}\partial_{k_a}H(k)\tau_z
\end{align}
where $a$, $b$ denote directions, and $\tau_i$ is a Pauli matrix acting on the Nambu space. 
With these velocity operators and the band structure,
we study linear and nonlinear conductivities $\sigma^{(1)}(\omega)$ and $\sigma^{(2)}(\omega)$.
In two-band systems with $\mathcal{T}$ symmetry, the linear conductivity~\cite{mahanbook} and nonlinear conductivity via shift current~\cite{sipe00prb,cook17natcomm} are given by
\begin{align}
    \sigma^{(1),ab}(\omega)
    &=
    \sum_{i,j}\int\frac{dk}{\omega}v^a_{ij}v^b_{ji}\delta(\omega-E_{ij}),
    \label{eq:lin1bsigma}
\end{align}
and
\begin{align}
    \sigma^{(2),abc}_{shift}(\omega)
    &=
    \sum_{i,j}\int\frac{dk}{\omega^2}f_{ij}\Im\left(v^a_{ij}w^{bc}_{ji}\right)\delta(\omega-E_{ij}),
    \label{eq:o21bsigma}
\end{align}
respectively.
Here, we set $e=1, \hbar=1$ for simplicity.
The full expressions for $\sigma^{(1)}(\omega)$ and $\sigma^{(2)}(\omega)$ for general systems are in the Supplemental Material~\ref{appsigmas}. 
Shift current is a dc current induced by light irradiation in $\I$-broken systems, which originates from nonzero polarization of photoexcited electron-hole pairs \cite{sipe00prb,rappe12,cook17natcomm,morimoto16,patankar18}.
There exist another mechanism for photocurrent, called injection current,
which arises from group velocity of photoexcited carriers and grows linearly in time.
Since injection current vanishes for linearly polarized light in the presence of $\TR$ symmetry, we focus on shift current in this paper.
In the following analysis, unless specified otherwise, we assume zero temperature.


{\it Absence of optical transitions within a pair.--} Now we show that there are no optical transitions between particle-hole pair bands in a BdG Hamiltonian with $\TR$ and $\I$ symmetries. 
Specifically, we show that the velocity matrix element connecting particle-hole pair bands in the BdG Hamiltonian under these symmetries is identically zero.
This fact is sometimes known as the vanishing linear conductivity at the superconducting gap in clean superconductors with $\TR$ and $\I$ symmetries~\cite{mahanbook}. 
The absence of optical transitions indicates that SHG and shift current also vanish at the superconducting gap, which is natural because the presence of $\I$ symmetry forbids second-order nonlinear optical effects.

To show this, we only need to calculate $\sigma^{(2)}_{shift}(\omega)$ and $\sigma^{(2)}_{SHG}(\omega)$ between the two particle-hole pair bands. It turns out that we only need to calculate $\hat v^a(k)$ between these two bands, which gives zero. If we call these two states $1,2$, we can verify directly that the $v^{a}_{12}=0$ for particle-hole pair bands in BdG Hamiltonian when the superconductor preserves $\TR$ and $\I$, as detailed in the Supplemental Material \ref{apppf} for completeness.
We note that $\I$ breaking in the normal part of the Hamiltonian is not sufficient and the $\I$ breaking in the SC order parameter is necessary for $v^{a}_{12} \neq 0$. For example, if the gap function is $\I$-symmetric as $\Delta= \Delta_0 \mathbb{1}_N$ ($\mathbb{1}_N$: an identity matrix with the dimension $N$), the transition matrix element $v_{12}^a$ identically vanishes even when the normal part $H(k)$ breaks $\I$ symmetry. In a general setup, some $\I$ breaking in $\Delta$ is expected when the normal part breaks $\I$ symmetry. 

This shows that a superconductor within the BdG description (\eq{bdg}) has no optical resonance between particle-hole pair bands as long as $\I$ and $\TR$ symmetries are preserved. Once $\I$ symmetry is broken, there can be nonzero optical resonance at these pair bands including at superconducting gap.  We will demonstrate this optical resonance by explicit computation in several models and draw some general conclusions about its strength.

\begin{figure}
    \centering
    \begin{subfigure}[t]{.235\textwidth}
        \centering
        \caption{}\includegraphics[width=\linewidth]{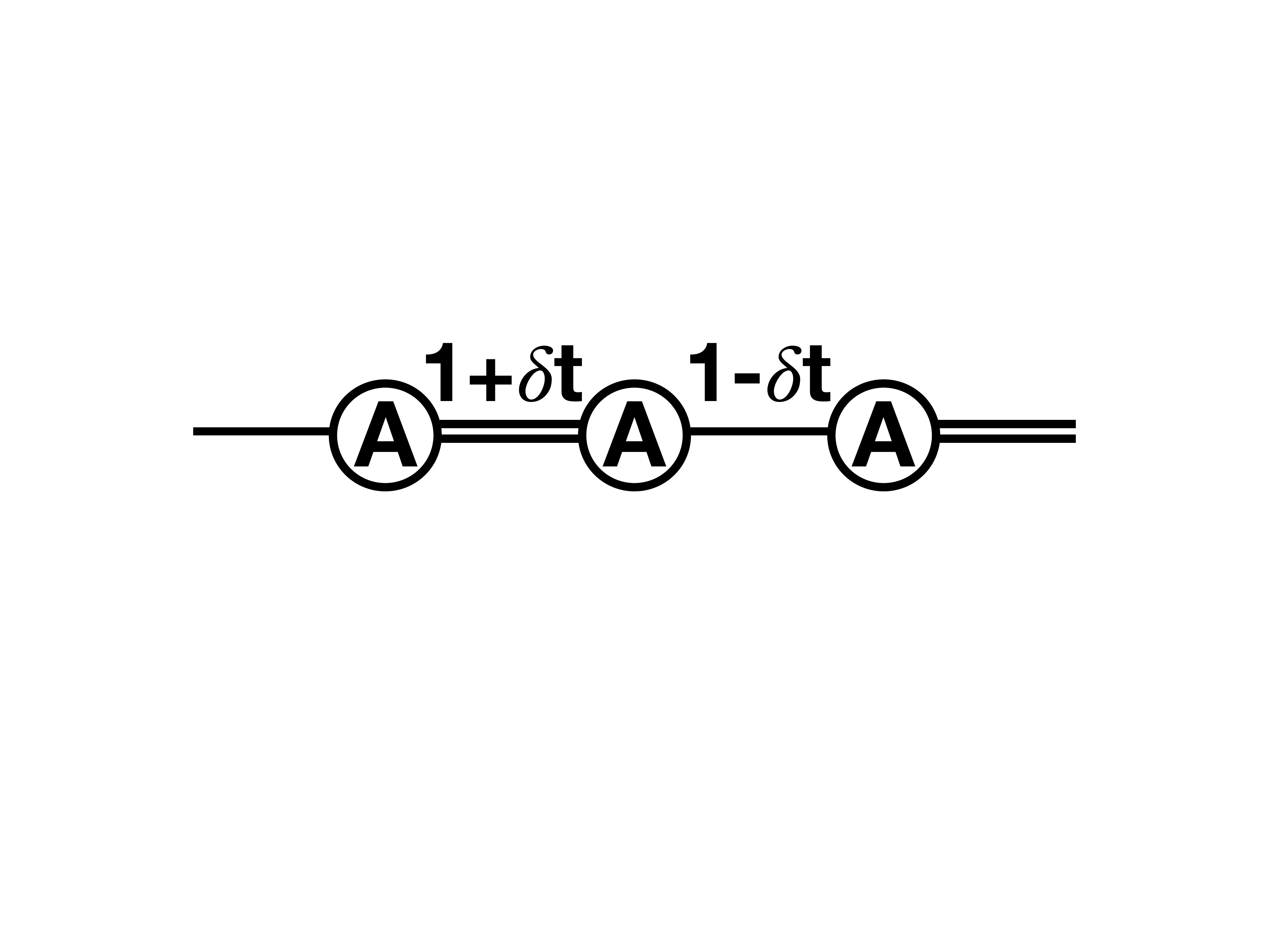}
        \label{fig:1D2Ba}
    \end{subfigure}
    \begin{subfigure}[t]{.23\textwidth}
        \centering
        \caption{}\includegraphics[width=\linewidth]{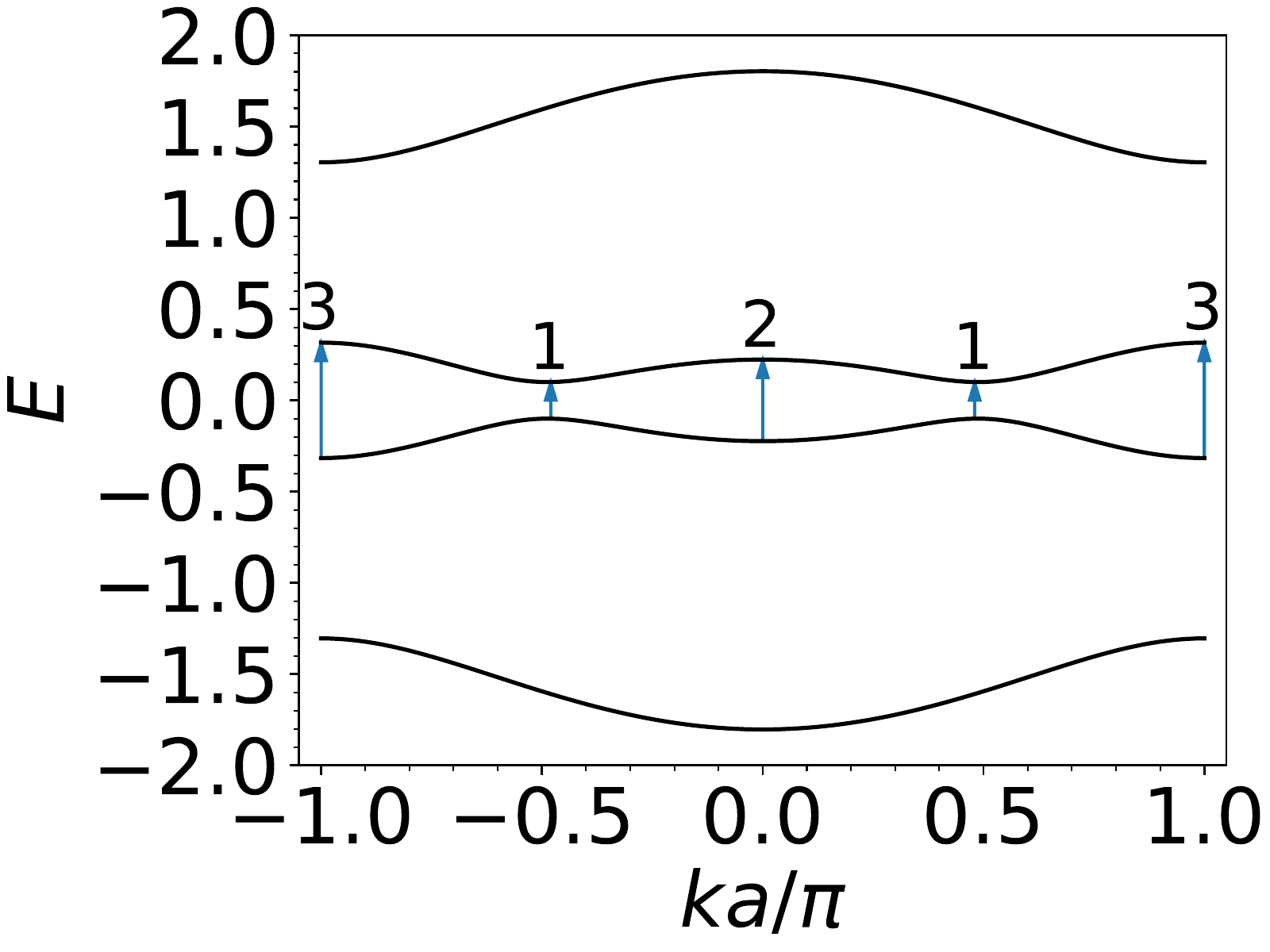}
        \label{fig:1D2Bb}
    \end{subfigure}
    \begin{subfigure}[t]{.22\textwidth}
        \centering
        \caption{}\includegraphics[width=\linewidth]{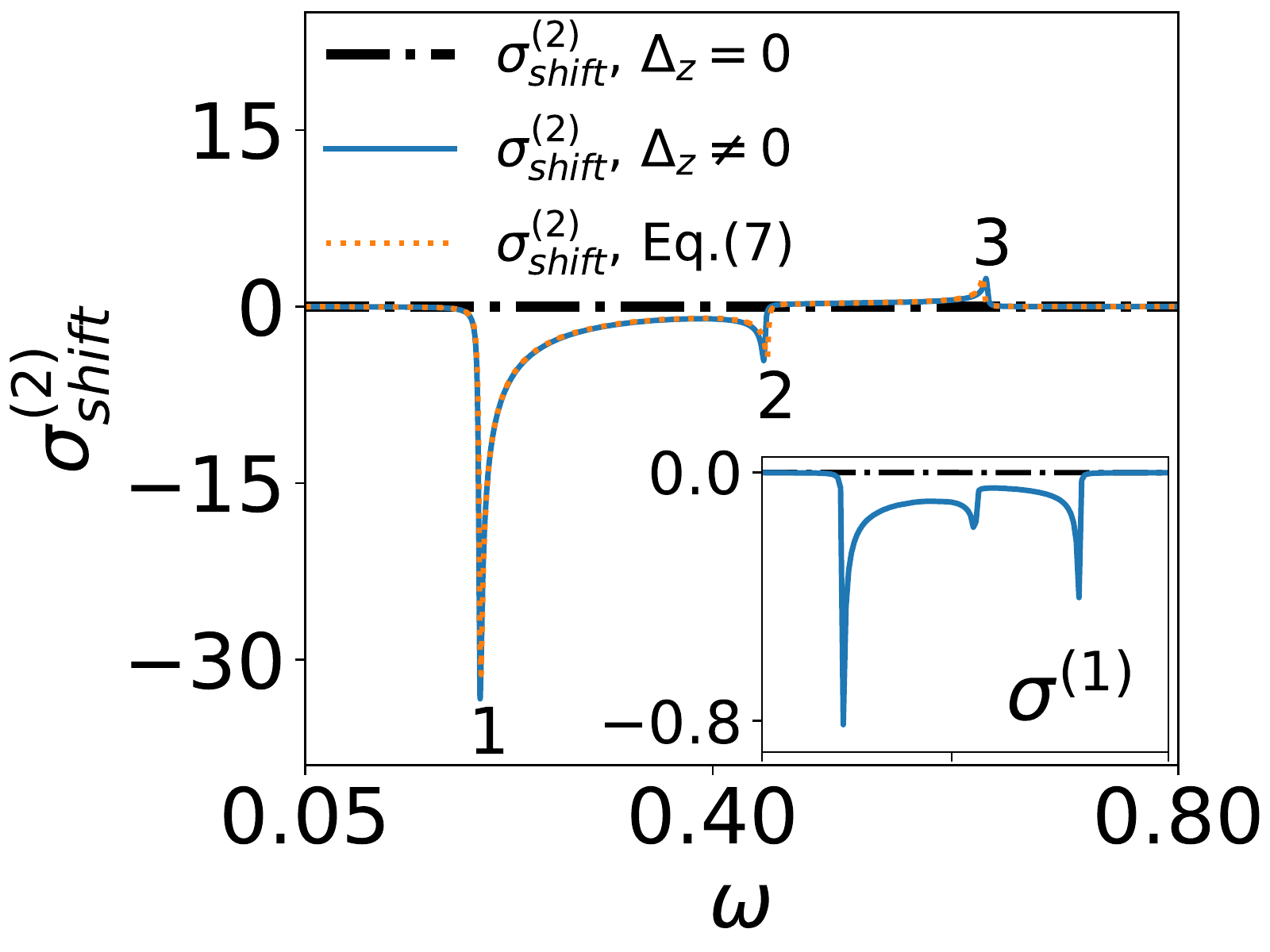}
        \label{fig:1D2Bc}
    \end{subfigure}
    \begin{subfigure}[t]{.24\textwidth}
        \centering
        \caption{}\includegraphics[width=\linewidth]{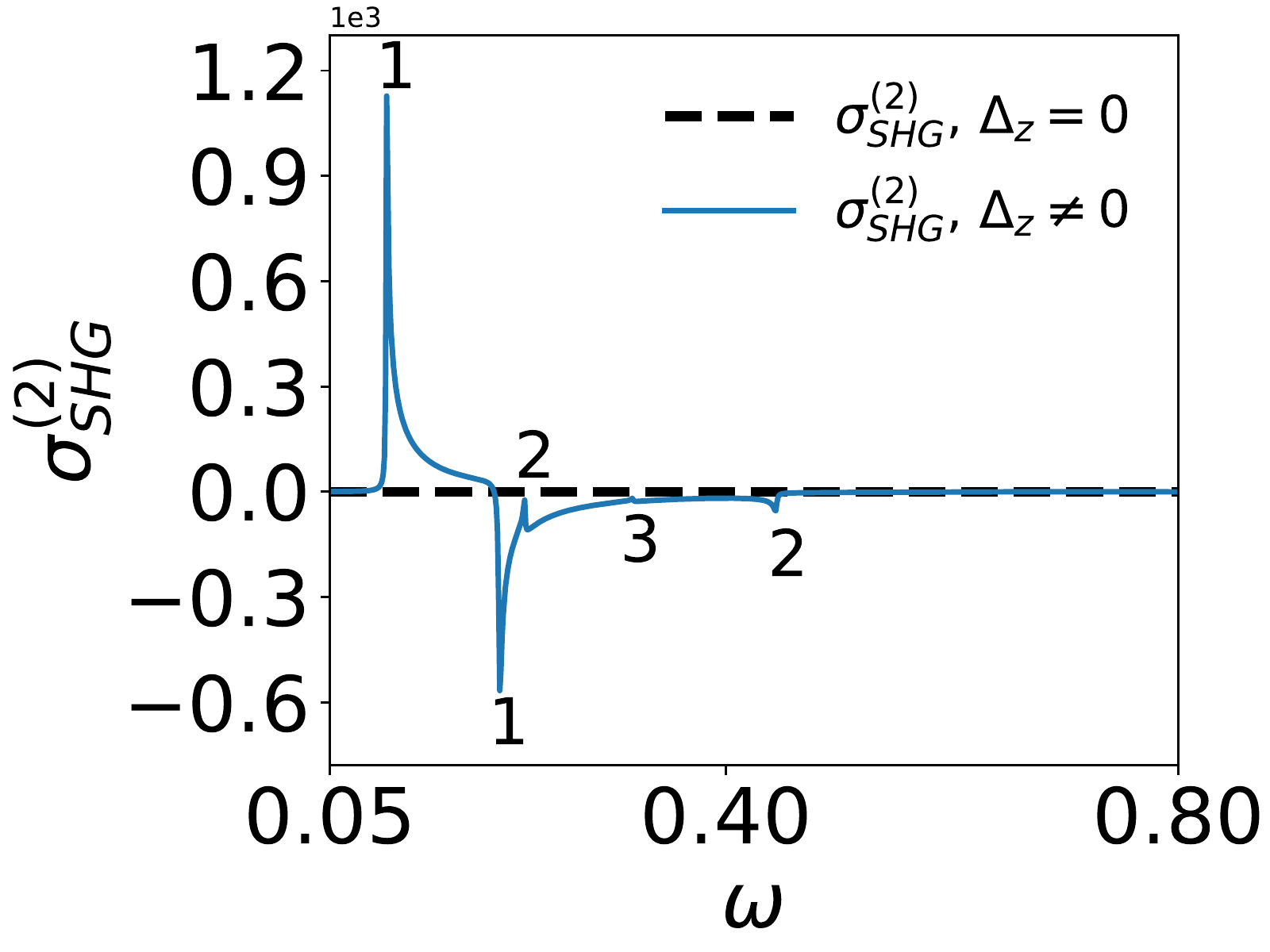}
        \label{fig:1D2Bd}
    \end{subfigure}
    \caption{
    Nonlinear optical responses of 1D, 2-band model (\eq{RM}) with $\I-$breaking SC gaps (\eq{1d2bgap}). 
    \sfig{1D2Ba} Schematic plot of 1D two band model. Different bonds ($-$ and $=$) indicate different hopping amplitudes.
    \sfig{1D2Bb} The band structure.
    \sfig{1D2Bc} $\sigma^{(2)}_{shift}$ with (blue solid curve) and without (black dash-dotted curve) $\I$-breaking, $\Order(\Delta_z)$ contribution (orange dotted curve) to $\sigma^{(2)}_{shift}$ with $\I$-breaking. The inset shows linear conductivity, where the $x-$axis of the inset is the same as those of the main panel.
    \sfig{1D2Bd} $\sigma^{(2)}_{SHG}$ with (blue solid curve) and without (black dash-dotted curve) $\I$-breaking.
    In \sfig{1D2Bc} and \sfig{1D2Bd}, we match the peak positions with their corresponding transitions in the band structure in upper right panel. We identify the transitions in between of the particle-hole pairs that are closest to Fermi surface (blue arrows with solid lines).
    }
    \label{fig:1D2B}
\end{figure}

{\it $\I-$breaking induced conductivities in a 2-band system.--} We study linear and nonlinear conductivities in a 1D chain with alternating hopping and an $\I-$breaking superconducting gap, which is the simplest model of superconductors with broken $\I-$symmetry while the normal (non-superconducting) part of the electronic structure preserves $\I$.  Specifically, we consider \eq{bdg} with the Hamiltonian
\begin{equation}
    H(k)=\cos{k}\sigma_x+\delta t\sin{k}\sigma_y-\mu,
    \label{eq:RM}
\end{equation}
where $\sigma_i$'s are the Pauli matrices, $\delta t,\mu\in\Re$. 
This model corresponds to the SSH model~\cite{ssh80}, or Rice-Mele model \cite{ricemele} with its on-site staggered potential $m$ being zero. 
We show a schematic plot of this model in \fig{1D2B}-(a).

We introduce the $\I$-breaking effect of $\Delta$ by adding a small symmetry breaking term, $\Delta_z\sigma_z$:
\begin{equation}
    \Delta=\Delta_0\sigma_0+\Delta_z\sigma_z,
    \label{eq:1d2bgap}
\end{equation}
where $\Delta_0,\Delta_z\in\Re$. 

We numerically study $\sigma^{(2)}_{shift}$ and $\sigma^{(2)}_{SHG}$ of this model.
We show the band structure and optical conductivity for this model in \fig{1D2B}, panels (b)-(d). We show shift current and SHG with (solid lines) and without (dash-dotted lines) $\I$-breaking, which indeed shows signals at $\omega\simeq2\Delta_0$ (shift current and SHG) and $\omega\simeq\Delta_0,~2\Delta_0$ (SHG) in the presence of $\I$-breaking, i.e. $\Delta_z\neq 0$. We match the $\sigma^{(2)}$ peaks with their corresponding transitions in the band structure panel. We also show linear conductivity calculations in the inset of \fig{1D2B} panel (c). This indicates that $\I$-breaking of $\Delta$ also causes non-zero {\it linear} effect.

\begin{figure}
    \centering
    \includegraphics[scale=0.5]{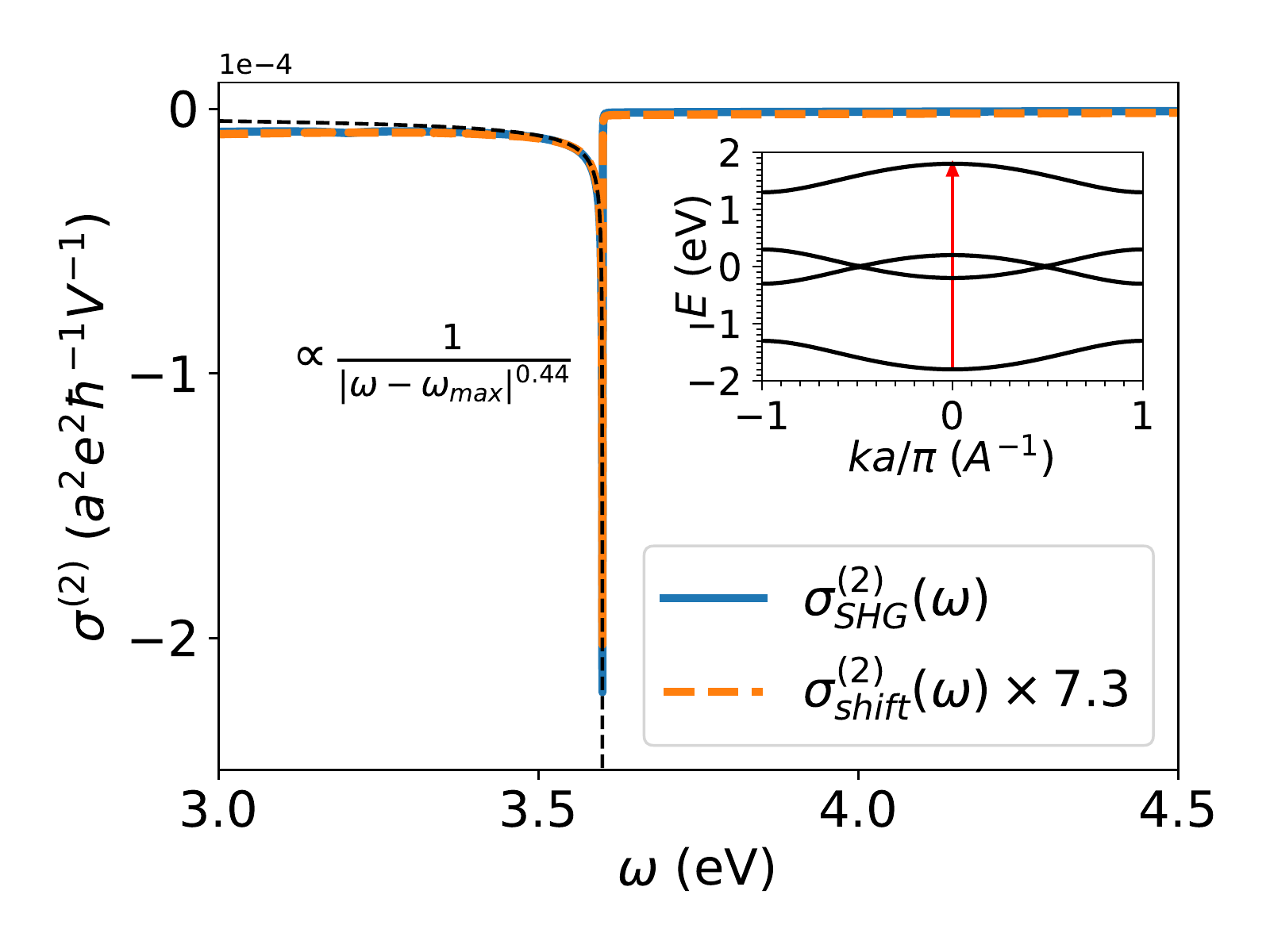}
    \caption{Frequency dependence of both shift current (orange, dashed curve) and SHG (blue, solid curve) at $\omega\simeq3.6~eV$ using realistic parameters: $t=1~eV$ (the hopping amplitude in front of $\sigma_x$), $\delta t=0.1~eV$, $\Delta=10~meV$ and $\Delta_z=2~meV$. Black dashed curve shows fitting. The peak corresponds to the energy difference of the two farthest bands at $k=0$. Inset shows the band structure, with red arrow indicating this transition. Note that the ``a'' in the unit of $\sigma^{(2)}$ is lattice constant.}
    \label{fig:1D_RMSc_real_fit}
\end{figure}

Next, we study the frequency dependence of $\sigma^{(2)}_{shift}$ and $\sigma^{(2)}_{SHG}$ of calculations with realistic parameters in \fig{1D_RMSc_real_fit}. \fig{1D_RMSc_real_fit} indicates that one gets $\sigma^{(2)}$ signals not only at $\omega\sim\Delta$, but also at higher energies such as $3.6~eV$ as shown. The value of this frequency comes from the energy differences between the two bands that are the farthest away from Fermi level at $k=0$. 

We also indicate this transition with a red arrow in the inset of \fig{1D_RMSc_real_fit}. 
The exponent $0.44$ that we obtained for the scaling of the peak structure is close to $1/2$, and can be explained by the quadratic band dispersion at the band edge. 

\begin{figure}
    \centering
    \begin{subfigure}[t]{.23\textwidth}
        \centering
        \caption{}\includegraphics[width=\linewidth]{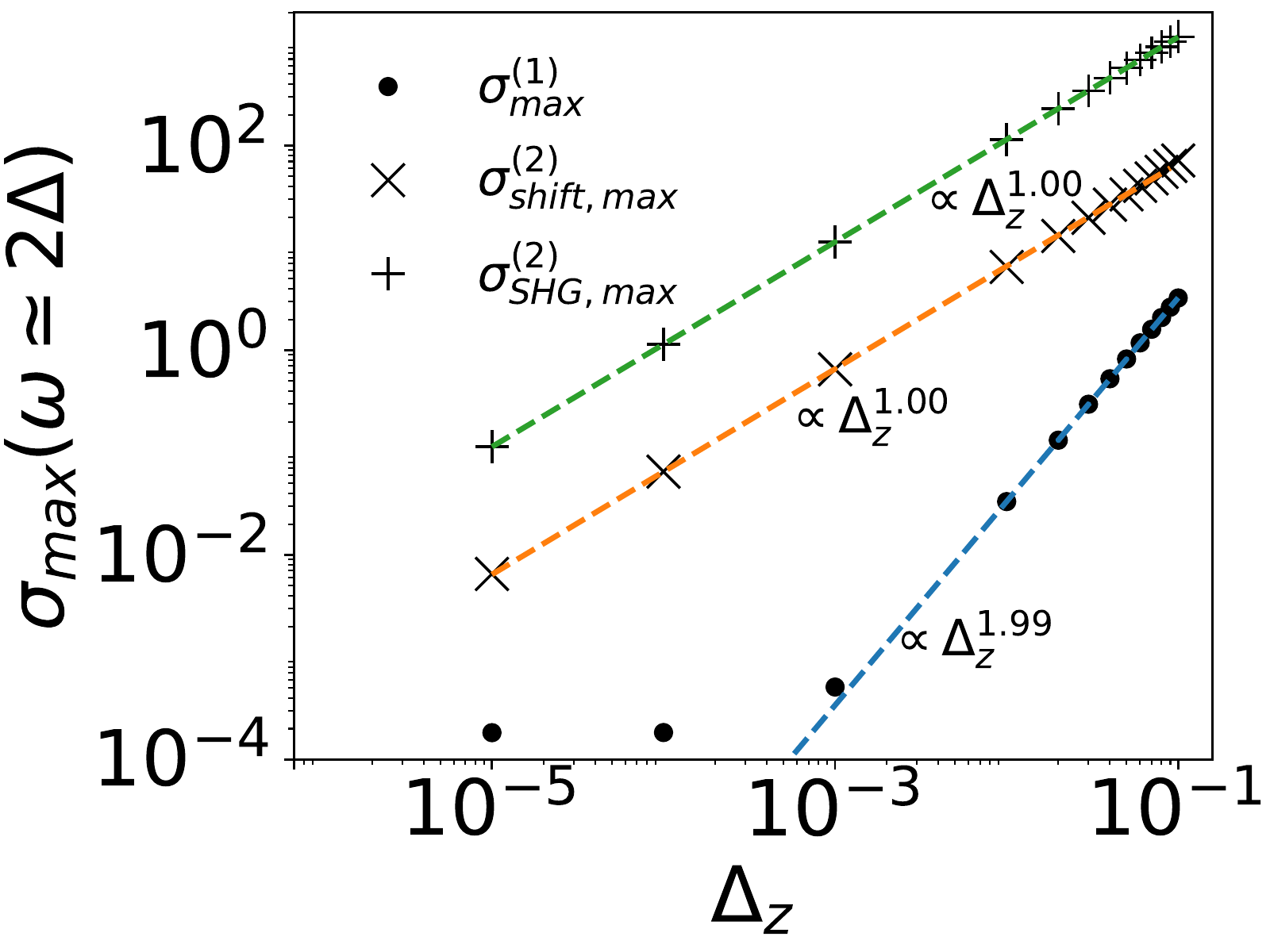}
        \label{fig:scalea}
    \end{subfigure}
    \begin{subfigure}[t]{.24\textwidth}
        \centering
        \caption{}\includegraphics[width=\linewidth]{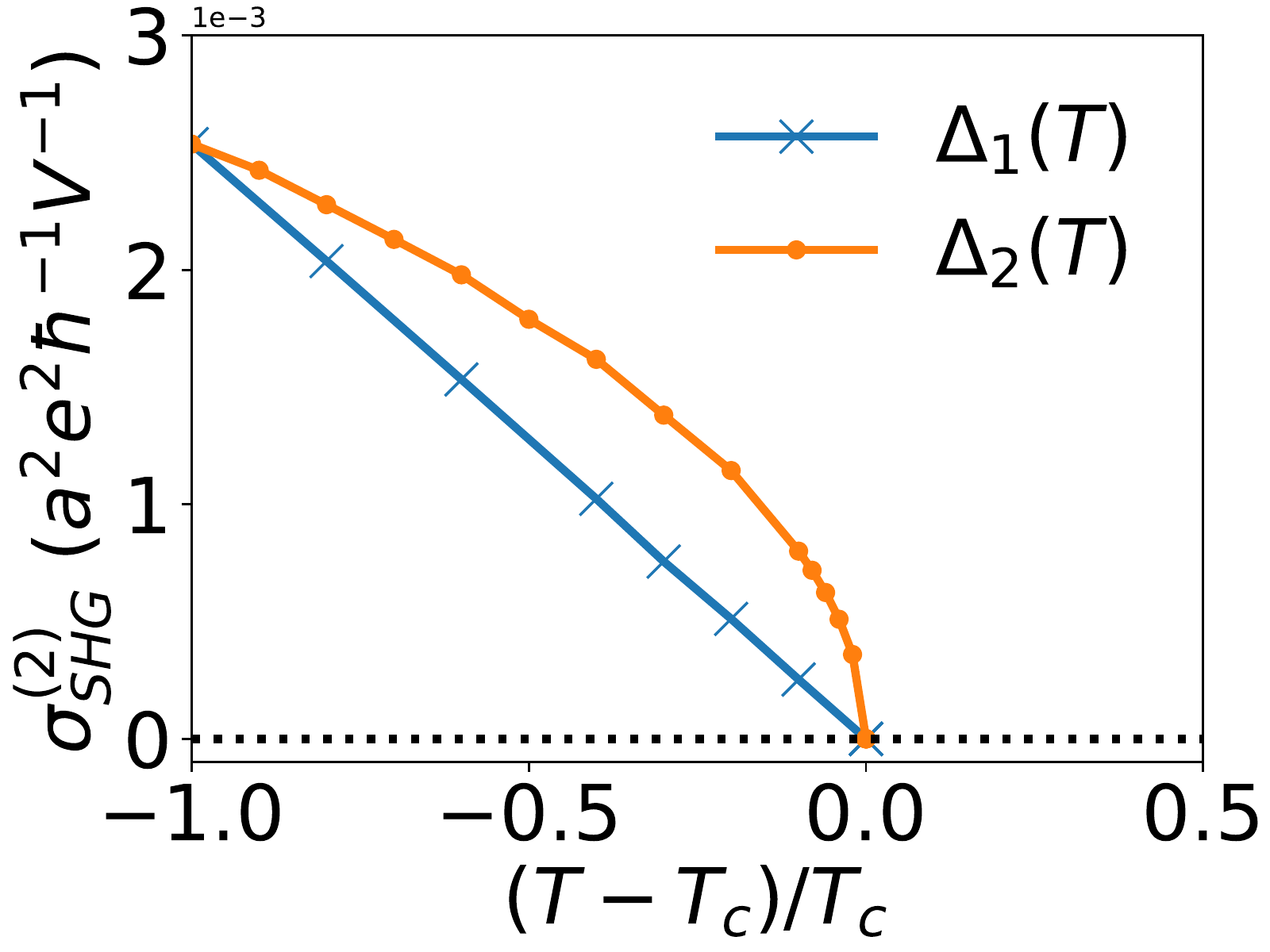}
        \label{fig:scaleb}
    \end{subfigure}
    \caption{
    \sfig{scalea} Scalings of local extrema at $\omega\simeq2\Delta$ of $\sigma^{(1)}$ (dots), $\sigma^{(2)}_{shift}$ (crosses) and $\sigma^{(2)}_{SHG}$ (pluses) with respect to $\I-$breaking SC order $\Delta_z$, fitted to lines in log-log scale. Parameters of the left panel: $\delta t=0.5$, $\mu=0.8$, $\Delta=0.1$. 
    \sfig{scaleb} Temperature dependences of SHG peaks at around $2~eV$ of our model, with $\Delta_1(T)$ (dashed curve with cross markers) and $\Delta_2(T)$ (solid curve with dotted markers). Parameters of the right panel: $\delta t=0.1~eV$, $\Delta_0=10~meV$, $r=0.2$ and $\Delta_z=2~meV$.
    }
    \label{fig:1D_RMSc_scale}
\end{figure}

Now, we examine how conductivities scale with $\Delta_z$ by examining the scaling between $\Delta_z$ and the local extrema of $\sigma$'s at $\omega\sim2\Delta$. \fig{1D_RMSc_scale}-(a) shows that such scaling is linear for $\sigma^{(2)}_{shift/SHG}$ and quadratic for $\sigma^{(1)}$ with respect to $\Delta_z$. 
This indicates that $\I$ symmetry breaking generally gives rise to a stronger signal in the nonlinear conductivity $\sigma^{(2)}$'s than that in $\sigma^{(1)}$. 
The stronger effect of $\I$ breaking on nonlinear conductivity can be understood by perturbative expansion with respect to $\Delta_z$. 
We obtain the lowest order effect of $\Delta_z$ by expanding $\hat{v}_{ij}$ to linear in $\Delta_z$, and expanding $\hat{w}_{ij}$ to constant.
(For details, see Supplemental Material \ref{apppert}).
This allows us to derive an analytic expression for $\sigma^{(2)}_{shift}$ for the transition between of the two low energy bands as
\begin{align}
    \sigma_{shift}^{(2)} \simeq
    \frac{\delta t\Delta\Delta_z}{4\mu}\int\frac{dk}{E_1(k)^3}\left[1- \frac{\delta t^2}{\mu\xi_k^3} \right]\delta(&\omega-2\abs{E_1(k)})\nonumber\\
    &+ \Order(\Delta_z^2),
    \label{eq:sigma2Odz}
\end{align}
where $\xi_k$ is the dispersion, $E_1(k)$ is the energy of the state just below Fermi surface.

We evaluate \eq{sigma2Odz} numerically and plot the result in \fig{1D2B}-(c) (orange, dotted curve), which lies almost on top of the numerical $\sigma^{(2)}_{shift}$ result. This shows that the largest, if not only, contribution to $\sigma^{(2)}_{shift}$ at energy close to $2\Delta$ is from the two lowest states of the system.

Now, we study how temperature affects SHG signals by incorporating the temperature dependence of the superconducting order parameter. 
We consider two scenarios for the $\I$-breaking SC order parameter:
\begin{enumerate}[label=\roman*]
    \item The {\it ratio} of $\I-$breaking is fixed below $T_c$ as
        \begin{equation}
            \Delta_1(T)=C_1(T_c-T)\Delta_0(\sigma_0+r\sigma_z);
            \label{eq:deltat1}
        \end{equation}
    \item The {\it amount} of $\I-$breaking is fixed below $T_c$ as
        \begin{equation}
            \Delta_2(T)=C_2(T_c-T)\Delta_0\sigma_0+\Delta_z\sigma_z.
            \label{eq:deltat2}
        \end{equation}
\end{enumerate}

We study the effect of these two scenarios by examining the temperature dependence of SHG signal at around $2eV$ of our model (\eq{RM}), with 
$
C_1(T_c-T)=C_2(T_c-T)=\sqrt{1-T/T_c}.
$
The optical measurement on cuprate superconductors found an SHG signal below $T_c$~\cite{zhao16natphys},
which shows a rapid signal drop as $T$ approaches the superconducing transition temperature $T_c$.
Comparing the experimental signal with our result shown in \fig{1D_RMSc_scale}-(b), we find that the experimental signal in Ref.\cite{zhao16natphys} may be a result of scenario (ii) for the inversion-breaking gap function (\eq{deltat2}). 

\begin{figure}
    \centering
    \begin{subfigure}[t]{.06\textwidth}
        \centering
        \caption{}\includegraphics[angle=90,width=.4\linewidth]{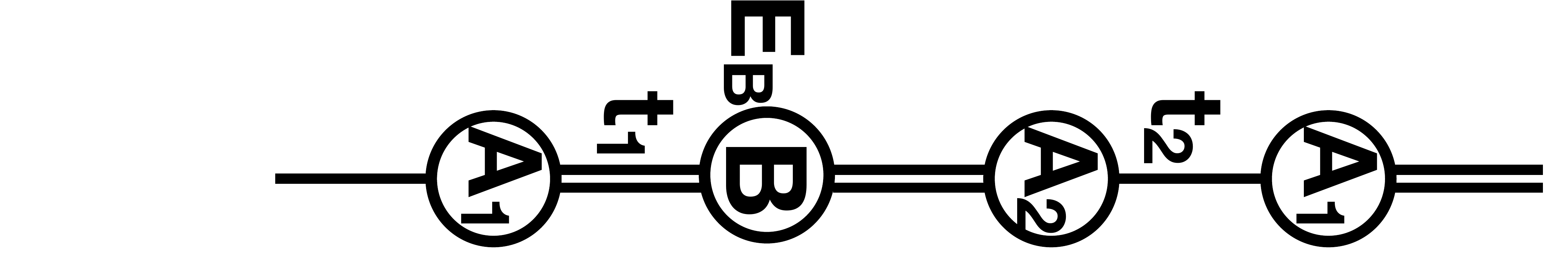}
        \label{fig:1d3ba}
    \end{subfigure}
    \begin{subfigure}[t]{.2\textwidth}
        \centering
        \caption{}\includegraphics[width=\linewidth]{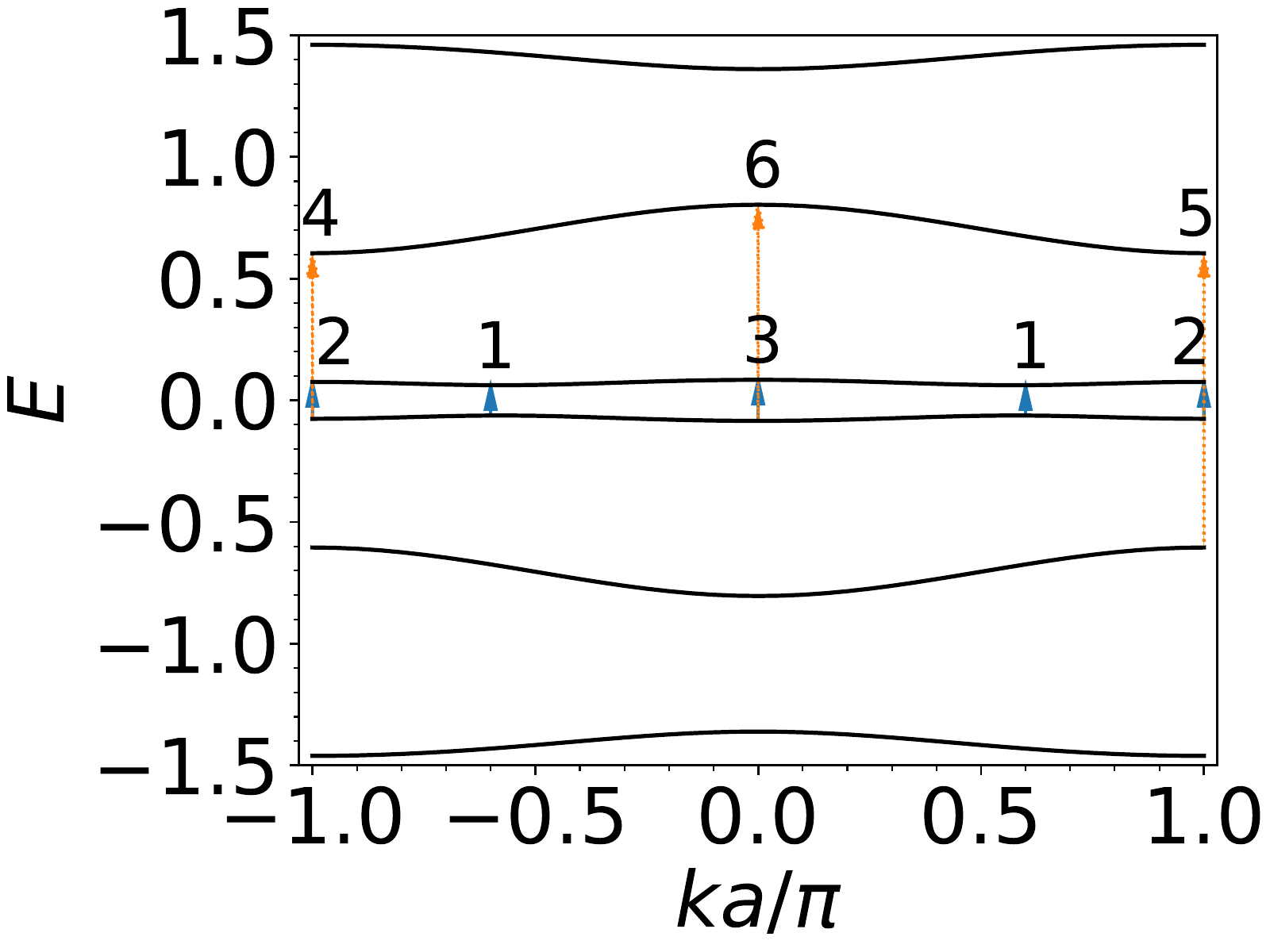}
        \label{fig:1d3bb}
    \end{subfigure}
    \begin{subfigure}[t]{.21\textwidth}
        \centering
        \caption{}\includegraphics[width=\linewidth]{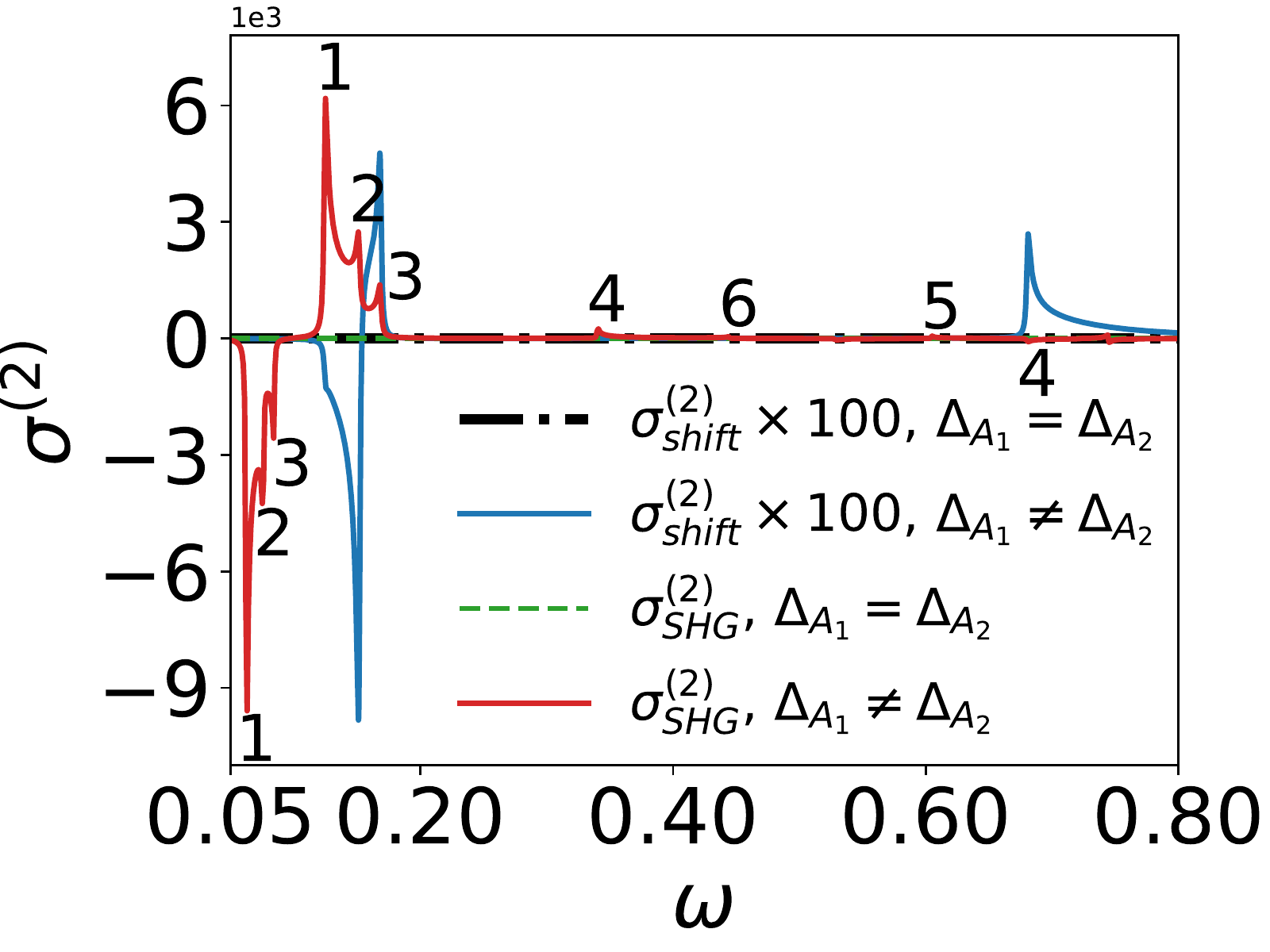}
        \label{fig:1d3bc}
    \end{subfigure}
    \caption{
    Nonlinear optical responses of 1D, 3-band model (see Supplemental Material. \ref{app1D3B}), with $\I-$breaking SC gaps (\eq{1d2bgap}). 
    \sfig{1d3ba} Schematic plot of 1D, 3-band model. Different bonds ($-$ and $=$) indicate different hopping amplitudes, and different particle species ($A$ and $B$) indicate different on-site potentials.
    \sfig{1d3bb} The band structure.
    \sfig{1d3bc} $\sigma^{(2)}_{shift}$ (black and blue curves) and $\sigma^{(2)}_{SHG}$ (green and red curves) with (solid curve) and without (dash-dotted curve) $\I$-breaking. 
    The SHG peak positions are associated with their corresponding transitions in \sfig{1d3bb}.
    }
    \label{fig:1D3B}
\end{figure}

{\it $\I-$breaking induced conductivities in 3-band system.--} To demonstrate that the nonvanishing nonlinear response generally appears in inversion-broken superconductors beyond the simplest two-band model, we briefly consider a 3-band system~\cite{fregoso17prb}.
The model consists of three sites connected with hoppings such that both bond and site centered $\I$ symmetry are preserved in the normal state. 
The detail of this model is in Supplemental Material~\ref{app1D3B}.
We also show a schematic plot of this model in \fig{1D3B}-(a).

The $\I$ symmetry may be broken by setting the superconducting paring via:
\begin{equation}
    \Delta=\begin{pmatrix}
                \Delta_{A_1} & 0 & 0\\
                0 & \Delta_{B} & 0\\
                0 & 0 & \Delta_{A_2}
            \end{pmatrix},
    \label{eq:1d3bgap}
\end{equation}
with $\Delta_{A_1}\neq\Delta_{A_2}$. Note that, for this system, the electronic part preserves {\it both} bond {\it and} site inversions, and all the $\I-$breakings are caused by $\Delta$. We show band structure, together with shift current and SHG of this system in \fig{1D3B}-(b)(c). Our result shows that, as we break the $\I$ symmetry, the optical transitions are indeed induced.

{\it Conclusion.--}In this work, we have shown the importance of $\I$-breaking superconductivity in both linear and non-linear optical effects in 1D superconducting system and a 2D cuprate minimal model (see Supplemental Material~\ref{app2D}). Our results indicate that not only the underlying symmetry of electronic structure, but also that of superconductivity itself, may give rise to nontrivial optical effects.
The scaling with respect to $\I-$breaking order parameter indicates that the distinctive optical effects are stronger in the second-order response $\sigma^{(2)}$ than in linear conductivity $\sigma^{(1)}$.  By computing non-linear optical effects with realistic parameters, we show that such effects can be detected in an experimentally accessible energy region.

Our analysis also shows a possible explanation of the observation in Ref.~\cite{zhao16natphys} that the temperature dependence of experimental SHG signals, even at an energy that may be well above than superconducting gap, may come from $\I$-breaking of the superconducting pairing. Our results also are consistent with the observation that $\sigma^{(1)}$ signals are weaker in such systems, especially when $\I$-breaking is small.

The search for new unconventional superconductors is a major goal of quantum condensed matter physics, but the determination of the symmetry breaking in the superconducting state remains challenging.  The ``gold standard'' of phase-sensitive Josephson tunneling to determine order parameter symmetry~\cite{buckleyreview} has only been feasible for a small set of materials.  The results presented here show that the signatures of inversion-symmetry breaking in nonlinear optical quantities are strong (relative to linear signatures) and persistent over a range of temperatures, and we hope that these observations will aid in the characterization of new superconductors.

\textit{Acknowledgements. ---}
We thank J. Orenstein, N. Nagaosa and D. Parker for fruitful discussions. Some of the optical conductivities results are computed using a numerical integration wrapper \cite{cubaturewrapper}. 
The authors were supported by the  U.S. Department of Energy (DOE), Office of Science, Basic Energy Sciences (BES), under Contract No. AC02-05CH11231 within the Ultrafast Materials Science Program (KC2203).

\bibliography{cmp}

\newpage
\widetext
\clearpage
\begin{center}
\textbf{\large Supplemental Materials}
\end{center}
\label{supp}
\setcounter{equation}{0}
\setcounter{figure}{0}
\setcounter{table}{0}
\setcounter{page}{1}
\makeatletter
\renewcommand{\theequation}{S\arabic{equation}}
\renewcommand{\thefigure}{S\arabic{figure}}

\section{Linear conductivity, shift current and second harmonic generation}\label{appsigmas}
Here we show how we compute conductivities. Following small $eA$ expansion of $H(k+eA)$ and set $e=\hbar=1$, we have linear conductivity \cite{mahanbook}:
\begin{equation}
    \sigma^{(1),ab}(\omega)\propto\sum_{i,j}\int\frac{dk}{\omega}f_{ji}v^a_{ij}v^b_{ji}\delta(\omega-E_{ij}),
    \label{eq:linsigma}
\end{equation}
shift current \cite{sipe00prb,cook17natcomm}:
\begin{align}
    \Re\left[\sigma^{(2),abb}_{shift}(\omega)\right]\propto-\sum_{ij}\int\frac{dk}{\omega^2}f_{ij}\Im\Bigg\{v_{ij}^b\bigg[\frac{v_{ji}^b\Delta_{ji}^a+v_{ji}^a\Delta_{ji}^b}{E_{ji}}-w_{ji}^{ba}+\sum_{l\neq i,j}\bigg(\frac{v_{jl}^bv_{li}^a}{E_{li}}-\frac{v_{jl}^av_{li}^b}{E_{jl}}\bigg)\bigg]\Bigg\}\delta(\omega-E_{ji}).
    \label{eq:o2sigma}
\end{align}
and second harmonic generation (SHG) \cite{yang17arXiv}:
\begin{align}
    \Re\left[\sigma^{(2),abc}_{shg}(\omega,2\omega)\right]\propto-\sum_{ilj}\int\frac{dk}{\omega^2} \Big\{I_{1,ijl}^{abc}f_{ij}\delta(2\omega-E_{ji})+I_{2,ij}^{abc}f_{ij}\delta(\omega-E_{ji})+I_{3,ijl}^{abc}\left[f_{il}\delta(\omega-E_{li})-f_{jl}\delta(\omega-E_{jl})\right]\Big\},
\end{align}
where 
\begin{align}
    I^{abc}_{1,ijl}&=\Im\left[v^a_{ij}w^{bc}_{ji}+\frac{2v^a_{ij}(v^b_{jl}v^c_{li}+v^c_{jl}v^b_{li})}{E_{il}+E_{jl}}\right],
    &I^{abc}_{2,ij}&=\Im\left(w^{ab}_{ij}v^c_{ji}+w^{ac}_{ij}v^b_{ji}\right),
    &I^{abc}_{3,ijl}&=\Im\left[\frac{v^a_{ij}(v^b_{jl}v^c_{li}+v^c_{jl}v^b_{li})}{E_{li}+E_{lj}}\right],
\end{align}
and
\begin{align}
    v^a_{ij}&=[\hat v^a(k)]_{ij}=\braket{u_i(k)|\hat{v}^a(k)\tau_0|u_j(k)}, &w^{ab}_{ij}&=[\partial_{k_a}\hat{v}^b(k)]_{ij}=\braket{u_i(k)|\partial_{k_a}\hat{v}^b(k)\tau_3|u_j(k)},\nonumber\\
    E_{ij}&=E_{ij}(k)=E_{i}(k)-E_{j}(k), &f_{ij}&=f(E_i)-f(E_j),
    &\Delta_{ij}^a&=v^a_{ii}-v^a_{jj},
\end{align}
with $f(E)$ being the Fermi-Dirac distribution. 
We also take into account the level broadening of the system by introducing a constant $\gamma$, and rewrite
\begin{align}
    \delta(\omega-E)\to\Im\frac{1}{\omega-E-i\gamma}.
\end{align}

\section{Proof of $v_{12}=0$ under $\mathcal{T}$ and $\mathcal{I}$ symmetries}\label{apppf}
In this section, we show that the velocity matrix element $v_{12}$ for bands 1 and 2 that are particle-hole symmetric partners vanishes under time reversal $\mathcal{T}$ and inversion $\mathcal{I}$ symmetries. 

For BdG Hamiltonian $H(k)$, we can represent the particle hole symmetry $\PH$ and time reversal symmetry $\TR$ as
\begin{align}
\PH H(k) \PH^{-1} &= - H(-k), & \PH &= \sigma_y\tau_y \K, \\
\TR H(k) \TR^{-1} &= H(-k), & \TR &= i\sigma_y \K,
\end{align}
with $\sigma$ and $\tau$ are the Pauli matrices acting on spin and Nambu spaces, respectively, and $\K$ denotes complex conjugation\cite{ryu08prb}.
Combining these two indicates there exist particle-hole pair bands $\psi_1$ and $\psi_2$ for given $k$ as
\begin{align}
H(k) \psi_1(k) &= E(k) \psi_1(k), & \psi_1 = \vp{u}{v}, \\
H(k) \psi_2(k) &= - E(k) \psi_2(k), & \psi_2 = \vp{v}{-u},
\end{align}
where $u$ and $v$ are $2N$ dimensional vectors including spin degrees of freedom. The velocity matrix element is given by
\begin{align}
v_{12} = \langle \psi_1| \hat v \tau_0 | \psi_2 \rangle = \langle u| \hat v|v \rangle - \langle v|\hat v|u \rangle,
\label{eq: v12}
\end{align}
where $\hat v$ is the velocity operator in the normal state.
Using the inversion symmetry $\I$,
we consider the combined $\I\TR$ symmetry.
If we represent $\I\TR=U_{\I\TR}\mathcal{K}$ with some unitary $U_{\I\TR}$ that is closed in the particle/hole space, we obtain the symmetry constraints as
\begin{align}
U_{\I\TR} H^*(k) U_{\I\TR}^{-1} &= H(k), & U_{\I\TR} \hat v^*(k) U_{\I\TR}^{-1} &= \hat v(k), 
\end{align}
and 
\begin{align}
u(k) &= U_{\I\TR} u^*(k), & v(k) &=U_{\I\TR} v^*(k).
\end{align}

Using the relationship to the second term for $v_{12}$ in \eqref{eq: v12} gives
\begin{align}
\langle v|\hat v|u \rangle 
&= \langle u^*|\hat v^*|v^* \rangle 
= \langle u| U_{\I\TR} U_{\I\TR}^{-1} \hat v U_{\I\TR} U_{\I\TR}^{-1}|v \rangle 
\nonumber \\
&= \langle u|\hat v|v \rangle.
\end{align}

Thus the velocity matrix element identically vanishes, i.e., 
\begin{align}
v_{12} =0,
\end{align}
for the bands 1 and 2 that are particle-hole symmetric partners.

\section{Perturbative calculation of $\sigma^{(2)}_{shift}$}\label{apppert}
Here we provide a sketch of the perturbative calculation of $\sigma^{(2)}_{shift}$ at $\omega\simeq E_{12}$, in 1D, 2-band model. \eq{o2sigma} tells us that we need to compute $\hat v_{12}$, $\hat v_{11}$,$\hat v_{22}$, $\hat v_{13}$, $\hat v_{23}$, $\hat v_{14}$, $\hat v_{24}$, $\hat w_{12}$, where $3$ and $4$ states are another particle-hole pair states.

Since $\hat v_{12}=0+\Order(\Delta_z)$, we need to compute the terms inside $\{\cdots\}$ of \eq{o2sigma} to $\Order(1)$. We only need to compute: $\hat v_{12}$, $\hat v_{13}$, $\hat v_{23}$, $\hat v_{14}$, $\hat v_{24}$, $\hat w_{12}$.

We diagonalize our BdG equation to $\Order({\Delta_z})$ using Mathematica. The $\Order({\Delta_z})$ parts of the eigenvectors are only needed for computing $\hat v_{12}$.

Hence we get $\hat v_{12}$ to $\Order(\Delta_z)$, and the rest to $\Order(1)$ as follows:
\begin{equation}
    \hat v_{12}=2i\frac{N_1N_2VE_1\sin{(\phi+\psi)}}{\Delta\mu}\Delta_z,
\end{equation}
\begin{equation}
    \hat w_{12}=4N_1N_2\xi_k,
\end{equation}
\begin{align}
    \hat v_{13}=-2i&N_1N_3V\sin{(\phi+\psi)}\left[\frac{(E_1+\xi_k-\mu)(E_3-\xi_k-\mu)}{\Delta^2}\right],
\end{align}
\begin{align}
    \hat v_{23}=\hat v_{13} &\text{\quad with \quad} 1\to2,
    &\hat v_{14}=\hat v_{13} &\text{\quad with \quad} 3\to4,
    &\hat v_{24}=\hat v_{13} &\text{\quad with \quad} 1\to2 \text{ and } 3\to4,
\end{align}
where
\begin{align}
    E_1&=-E_2=-\sqrt{(\xi_k-\mu)^2+\Delta^2}, &E_3&=-E_4=-\sqrt{(\xi_k+\mu)^2+\Delta^2},\nonumber\\
    N_{1/2}^{-2}&=\frac{2}{\Delta^2}\left[(E_{1/2}+\xi_k-\mu)^2+\Delta^2\right], &N_{3/4}^{-1}&=\frac{2}{\Delta^2}\left[(E_{3/4}-\xi_k-\mu)^2+\Delta^2\right],\nonumber\\
    \xi_k&=\sqrt{\cos^2k+\delta t^2\sin^2k},
    &V&=\sqrt{\sin^2k+\delta t^2\cos^2k},\nonumber\\
    e^{i\phi}&=\frac{\cos k+i\delta t\sin k}{\xi_k},
    &e^{i\psi}&=\frac{\sin k+i\delta t\cos k}{V}.
\end{align}

\section{1D, 3-band model}\label{app1D3B}
Here we specify the 3-band model\cite{fregoso17prb} that we consider in our analysis.

The model we consider is
\begin{equation}
    H_{3B}=\sum_jt_jc_j^\dagger c_{j+1} + H.c.,
\end{equation}
where $t_j's$ take three different values $t_1$, $t_2$ and $t_3$, then we may see this system having three sites in a unit cell. By setting $t_1=t_3$, and assuming the center cite of the unit cell having a different orbital from the rest, the Bloch Hamiltonian becomes:
\begin{equation}
    H_{3B}(k)=\begin{pmatrix}
                0 & t_1e^{ik}& t_2e^{-ik}\\
                t_1e^{-ik} & E_B & t_1e^{ik}\\
                t_2e^{ik} & t_1e^{-ik}& 0 
              \end{pmatrix}.
\end{equation}
The above hamiltonian preserves both bond and site inversion. Assuming spin SU$(2)$ (since we consider spin-singlet pairing for simplicity) and $\TR$ symmetry, our BdG hamiltonian becomes:
\begin{equation}
    H_{BdG}=
    \hp{H(k,\up)-\mu}{\Delta}{\Delta^*}{-H(k,\down)+\mu}.
\end{equation}

\section{A 2D Minimal Model}\label{app2D}
We here show the effect of $\I$-breaking in a 2D, d-wave superconducting minimal model on a square lattice.

For BdG hamiltonian in 2D system, we simply change $k$ in \eq{bdg} into $\vec k$. Then we consider the electronic structure in our system by coupling 1D Rice-Mel\'e chains into a 2D sheet, i.e.
\begin{equation}
    H(\vec k)=\cos k_x\sigma_0+\cos k_y\sigma_x+\delta t\sin k_y\sigma_y,
\end{equation}
which is $\I$-preserved. The bond alternation in this normal state could arise from some stripe order, for example, charge density waves. 

Then we turn on d-wave pairing \cite{tsuei00rmp}:
\begin{align}
    \Delta=\Delta_0(\cos k_x\sigma_0-\cos k_y\sigma_x),
\end{align}
and break its $\I$ by adding to $\Delta$ the term
\begin{align}
    \Delta'=\Delta_z\sigma_z.
\end{align}

\begin{figure}
    \centering
    \begin{subfigure}[t]{.23\textwidth}
        \centering
        \caption{}\includegraphics[width=\linewidth]{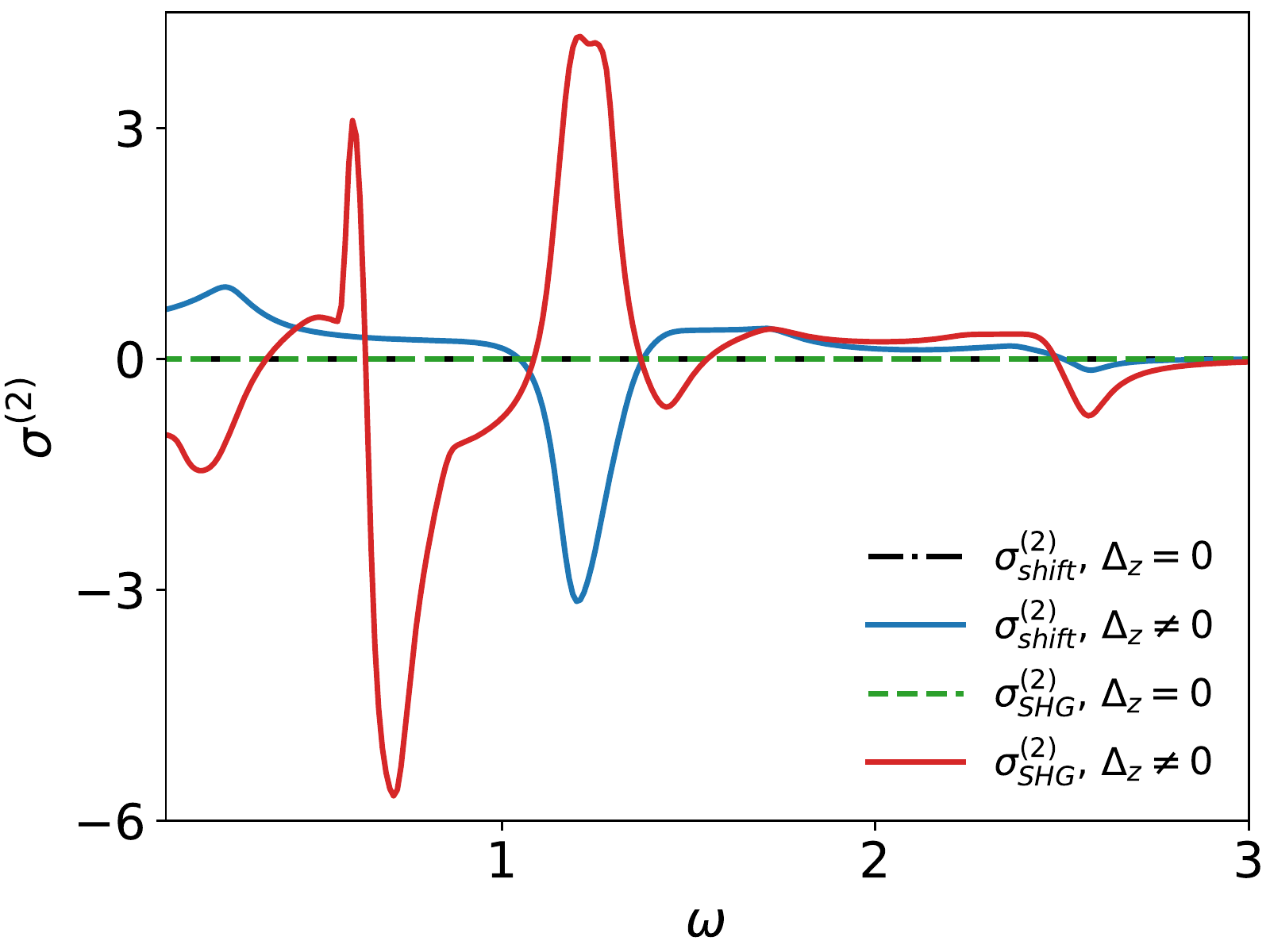}
        \label{fig:2da}
    \end{subfigure}
    \begin{subfigure}[t]{.23\textwidth}
        \centering
        \caption{}\includegraphics[width=\linewidth]{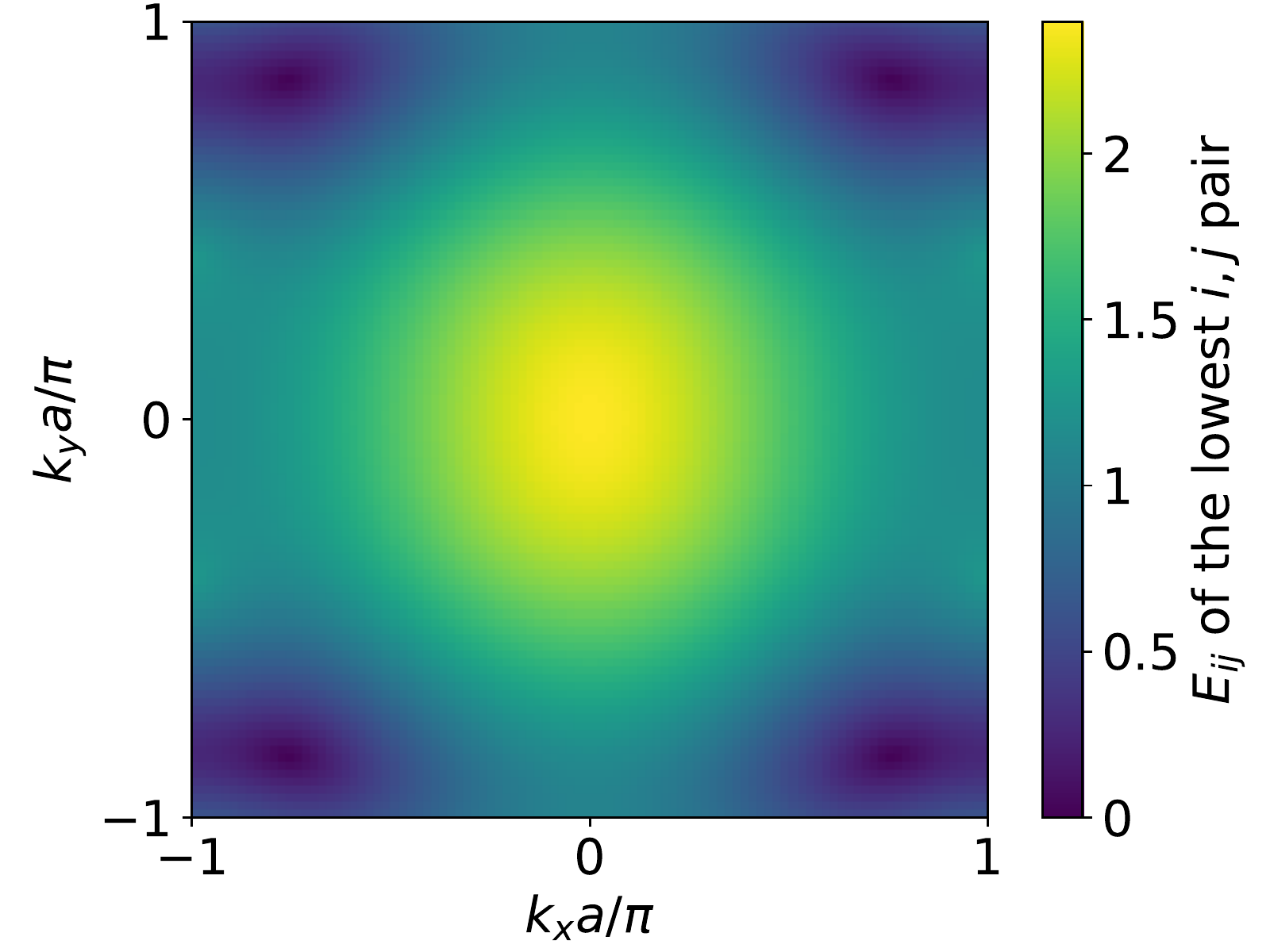}
        \label{fig:2db}
    \end{subfigure}
    \caption{2D minimal model calculation.
    \sfig{2da} $\sigma^{(2)}_{shift}$ and $\sigma^{(2)}_{SHG}$ with (solid curves) and without (dashed curves) $\I$-breaking. Parameters: $\delta t=0.5$, $\mu=0.8$, $\Delta_0=0.5$, $\Delta_z=0$ (dashed curves) and $\Delta_z=0.01$ (solid curves).
    \sfig{2db} band gap, $E_{ij}$, of the lowest pair of states.}
    \label{fig:2D_RMSc_o2}
\end{figure}

We compute $\sigma^{(2),yyy}_{shift}$ and show results in left panel of \fig{2D_RMSc_o2}. We see from the figure that, as expected, there is a peak at around $\omega\simeq2\Delta_0$ for $\I$-broken system, while no such signal for the $\I$-preserved one.

We also see peaks at $\omega<2\Delta_0$ and $\omega>2\Delta_0$ when $\Delta_z\neq0$. These may come from the Van Hove singularities\cite{vanhove53pr} in the density of states, induced by regions in the Brillouin zone where the two low-energy bands are parallel to each other, as can be seen from right panel of \fig{2D_RMSc_o2}.

\end{document}